\documentclass[preprint2]{aastex}

\def\be{\begin{equation}}
\def\ee{\end{equation}}

\begin{document}

\title{Ohmic Power of Ideal Pulsars} 

\author{Andrei Gruzinov}

\affil{CCPP, Physics Department, New York University, 4 Washington Place, New York, NY 10003}

\begin{abstract}

Ideal axisymmetric pulsar magnetosphere is calculated from the standard stationary force-free equation but with a new boundary condition at the equator. The new solution predicts Ohmic heating. About 50\% of the Poynting power is dissipated in the equatorial current layer outside the light cylinder, with about 10\% dissipated between 1 and 1.5 light cylinder radii. The Ohmic heat presumably goes into radiation, pair production, and acceleration of charges -- in an unknown proportion. 

~~

~~

\end{abstract}

\section{Introduction} 
We have shown that ideal pulsars calculated in the force-free limit of Strong-Field Electrodynamics (SFE) dissipate a large fraction of the Poynting flux in the singular current layer outside the light cylinder (Gruzinov 2011). This result -- finite damping in an ideal system -- is not really that unusual. Burgers equation, for instance, with viscosity $+0$, dissipates finite energy in infinitely thin shocks.

\begin{figure}
\plotone{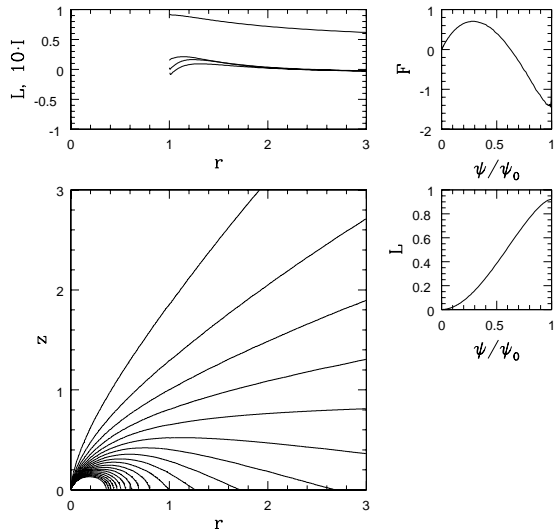}
\caption{Everything is in pulsar units. $r_s=0.25$. {\it Lower Left:} Isolines of $\psi$, integer multiples of $0.1\psi _0$, $\psi _0=1.44$. {\it Upper Left:} Poynting flux $L$ and the field invariant $I\equiv B^2-E^2$ at equator  vs $r$. Multiple curves for $I$ are different discrete approximations -- a rough rendering of numerical accuracy. {\it Upper Right:} $F$. {\it Lower Right:} Poynting flux $L$ through the light cylinder on the field lines with $\psi (1, z) < \psi$.} 
\end{figure}

The standard axisymmetric pulsar magnetosphere features a nearly head-on\footnote{$154.6^{\circ }$, Gruzinov (2005)} collision of Poynting fluxes right outside the light cylinder. It is to be expected, although merely by common sense, that such a collision should be accompanied by damping.

Here we show that our SFE solution also obtains from the standard force-free magnetosphere equation of Scharlemant \& Wagoner (1973), if one uses the ``correct'' boundary condition at the equatorial current layer.

We propose that the  ``correct'' boundary condition at the singular current layer (which now exists only outside the light cylinder) is 
\be \label{bc}
B^2-E^2=0.
\ee
This condition is Lorentz invariant, comes up in the SFE simulations \footnote{See Fig.5 of Gruzinov (2008).}, and has a clear physical meaning (at equator, the field becomes electric-like in order to drive large current). 

We cannot be sure that our proposal works, until one justifies the full SFE, or just eq.(\ref{bc}), microscopically. But conversion of 50\% of the Poynting flux into the Ohmic power (radiation, electron-positron pairs) occurring close to the light cylinder must have consequences for the pulsar phenomenology, and needs to be studied.

In \S2 we derive the pulsar magnetosphere equation and explain how Contopoulos, Kazanas \& Fendt (1999) solve it. In \S3 we put together all the equations which are needed to calculate the pulsar magnetosphere. In \S4 we describe the numerical solution and the corresponding physics results.

\section{Ideal pulsar magnetosphere} 

Goldreich \& Julian (1969) proposed that neutron star magnetospheres obey the force-free condition
\be \label{ff}
\rho {\bf E} + {\bf j}\times {\bf B}=0.
\ee
Surprisingly, it turns out that this simple equations allows a full calculation of the pulsar magnetosphere (Scharlemant \& Wagoner 1973, Contopoulos, Kazanas \& Fendt 1999, Gruzinov 2005, Spitkovsky 2006).

For the stationary axisymmetric case, the calculation is as follows. Using axisymmetry and stationarity, in cylindrical coordinates $(r,\theta ,z)$, we represent the fields by the three scalars $\phi$, $\psi$, and $A$, which depend on $r$ and $z$ but not on $\theta$:
\be \label{stat}
{\bf E} =-\nabla \phi,~~ {\bf B}={1\over r}(-\psi _z,A,\psi _r),
\ee
where the subscripts denote the partial derivatives. 

We plug (\ref{stat}) into (\ref{ff}) and use $\rho =\nabla \cdot {\bf E}$ and ${\bf j}=\nabla \times {\bf B}$. We also use the boundary conditions at the surface of the star -- the continuity of the normal component of the magnetic field and the tangential component of the electric field. We use the pulsar units 
\be 
\mu = \Omega =c=1,
\ee
where $\mu$ is the magnetic dipole moment of the star. It is assumed that the magnetic field is a pure dipole near the surface inside the star. The star is assumed to be a perfect conductor. $\Omega$ is the angular velocity of the star. 

We get 
\be 
\phi =\psi, ~~ A=A(\psi),
\ee
where $A$ is an arbitrary function of $\psi$, and we also get the ``Grad-Shafranov-like'' pulsar magnetosphere equation for $\psi$
\be \label{pme}
(1-r^2)\Delta \psi-{2\over r}\psi _r+F(\psi )=0.
\ee
Here $\Delta \equiv \nabla ^2$, and $F\equiv AA'$, where the prime denotes the $\psi$-derivative. The pulsar magnetosphere equation (\ref{pme}) is solved outside the star
\be 
r^2+z^2>r_s^2,
\ee 
with the boundary condition at the surface of the star
\be 
\psi ={r^2\over r_s^3}, ~~ r^2+z^2=r_s^2.
\ee 

The pulsar magnetosphere equation (\ref{pme}) contains $F$ --  an arbitrary function of $\psi$, and it is not clear how one should solve it. This was explained and done by Contopoulos, Kazanas \& Fendt (1999) (CKF).

The pulsar magnetosphere equation is elliptical both inside and outside the light cylinder, and can therefore be solved if some conditions are given at all boundaries and if $F$ is known. We first pick a trial $F$. 

Then inside the light cylinder, we have all the boundary conditions: (i) we know $\psi$ at the surface of the star, (ii) $\psi =0$ at $z=\pm \infty$, (iii) at $r=1$ the boundary condition is given by the pulsar magnetosphere equation itself, $\psi _r=F/2$. So we can find $\psi$ inside the light cylinder, say by the relaxation method using the variation principle.

Outside the light cylinder, CKF postulate the boundary condition at the equator,
\be 
\psi (r,0)=\psi (1,0), ~~r>1.
\ee 
Then, with some boundary conditions at infinity (we use $\psi =0$ at $z=\pm \infty$ and $\psi _r=0$ at $r=\infty$), and with the same boundary condition at the light cylinder, $\psi _r=F/2$ at $r=1$, one can solve the pulsar magnetosphere equation outside the light cylinder too.

For a generic $F$, this procedure gives a solution with $\psi (1-0,z)\neq \psi (1+0,z)$. But one might hope that there is a (unique?) $F$ which gives a smooth solution. CKF use a feedback procedure -- numerical adjustment of $F$ leading to a smooth light cylinder crossing.

In our case, the boundary condition at the equator outside the light cylinder is 
\be 
(r^2-1)(\nabla \psi )^2=A^2, ~~z=\pm 0,~r>1.
\ee 
At the light cylinder, $r=1$, this gives 
\be 
A(\psi _0)=0, ~~\psi _0\equiv \psi (1,0),
\ee 
meaning that there is no singular return current on the field line $\psi =\psi _0$. The only singularity is the equatorial current layer.

\section{Pulsar magnetosphere equation} 

In summary, we solve the following pulsar magnetosphere equation
\be \label{pme1}
(1-r^2)\Delta \psi-{2\over r}\psi _r+AA'=0,
\ee
\be \label{pme2}
\psi ={r^2\over r_s^3}, ~~ r^2+z^2=r_s^2,
\ee 
\be \label{pme3}
(r^2-1)(\nabla \psi )^2=A^2, ~~z=\pm 0,~r>1.
\ee 
The function $A(\psi)$ is (uniquely?) determined by the continuity at the light cylinder. The Poynting power is $L=\int d\psi A$. 

\section{Numerical Solution and Results} 

Our numerical procedure is as follows. We pick a trial function 
\be
g(r)>0,~~ g'(r)<0, ~~r>1, ~~g(1)=1.
\ee
Instead of the boundary condition (\ref{pme3}), we impose 
\be
\psi (r,0)=\psi _0g(r), ~~ r>1, ~~ \psi _0\equiv \psi (1,0).
\ee
We apply the CKF relaxation method to calculate $F(\psi )$ (ignoring the singular return current). Then we calculate the field invariant at the equator outside the light cylinder
\be 
I(r)={1\over r^2}(A^2+(\nabla \psi )^2)-(\nabla \psi )^2, ~~z=\pm 0,~r>1.
\ee 
We then {\it manually} chose $g(r)$, so as to make $I(r)$ as close to zero as we can at all $r>1$. 
It turns out, that 
\be
g(r)=0.52+{0.48\over r},
\ee
nullifies $I(r)$ to about the numerical accuracy \footnote{This numerical procedure, though formally correct, is methodologically inappropriate. In fact, we have ``cheated''. The function $g(r)$ has been read off the SFE time-dependent simulation (we then confirm that various other profiles $g(r)$ don't nullify $I(r)$ to the accuracy shown in Fig.1.). What one really wants is a CKF-type feedback loop, which would adjust $g(r)$ so as to nullify $I(r)$. Or, maybe, one can enforce the correct boundary condition (\ref{pme3}) throughout the $F$ relaxation. We were unable to develop a numerical scheme which would solve the problem (\ref{pme1} - \ref{pme3}) all by itself.}.

The resulting magnetosphere, the function $F(\psi )$, the Poynting flux at different radii, and the invariant $I(r)$ are shown in Fig.1. 

One gets the spin-down power (the Poynting flux at the stellar surface = the Poynting flux through the light cylinder):
\be
L_{\rm sd} \approx 0.9{\mu ^2 \Omega ^4\over c^3}.
\ee
The Ohmic power (the Poynting flux on the field lines that cross the equatorial current layer) is
\be
L_{\rm Ohm} \approx 0.5L_{\rm sd}.
\ee 
The Ohmic power between 1 and 1.5 light cylinder radii is 
\be
L_{\rm Ohm 1.5} \approx 0.1L_{\rm sd}.
\ee

\end{document}